\newcounter{myctr}
\def\myitem{\refstepcounter{myctr}\bibfont\noindent\ifnum\themyctr>9\else\phantom{0}\fi\hangindent17pt\themyctr.\enskip}

\documentclass{ws-ijqi}

\begin{document}

\markboth{E. A. Ivanchenko}{Entanglement dynamics in finite qudit
chain}

\catchline{}{}{}{}{}

\title{{ ENTANGLEMENT DYNAMICS IN  FINITE QUDIT CHAIN
\\IN  CONSISTENT MAGNETIC FIELD}}

\author{E. A. IVANCHENKO}

\address{National Science Center
\textquotedblleft{}Institute of Physics and
Technology\textquotedblright{}, Institute for Theoretical Physics,
Akademicheskaya str. 1, 61108 Kharkov, Ukraine \\
yevgeny@kipt.kharkov.ua}

\maketitle


\begin{abstract}
Based on the Liouville-von Neumann equation, we obtain a closed
system of equations for the description of a qutrit or coupled
qutrits in an arbitrary, time-dependent, external magnetic field.
The dependence of the dynamics on the initial states and the
magnetic field modulation is studied analytically and numerically.
We compare the relative entanglement measure's dynamics in
bi-qudits with permutation particle symmetry. We find the magnetic
field modulation which retains the entanglement in the system of
two coupled qutrits. Analytical formulae for the entanglement
measures in finite chains from 2 to 6 qutrits or 3 quartits are
presented.
\end{abstract}
   \section{Introduction}
Multi-level quantum systems are studied extensively, since they
have wide applications.  Some of the existing analytical
results\cite{Hioe} for spin 1 are derived in terms of a coherent
vector\cite{HioeEberly}. The class of exact solutions for a
three-level system is given in Ref.~\refcite{AMIshkhanyan}. The
application of coupled multi-level systems in quantum devices is
actively studied\cite{ZobovShauroErmilov}. The study of these
systems is topical in view of possible applications for useful
work in microscopic systems\cite{Scully}. Exact solutions for two
uncoupled qutrits interacting with the vacuum are obtained in
Ref.~\refcite{DerkaczJakobczyk}. For the case of  qutrits
interacting with a stochastic magnetic field, the exact solutions
are obtained in Ref.~\refcite{MazharAli}. The exact solutions for
coupled qudits in an alternating
magnetic field, to our knowledge, have not yet been found.\\
 The entanglement in multi-particle coupled systems is an important
resource for many problems in the quantum information science, but
its quantitative value computing  is difficult  because of
different types of entanglement. Multi-dimensional entangled
states are interesting both for the study of the foundations of
quantum mechanics and for the topicality of developing new
protocols for quantum communication. For example, it was shown
that for maximally entangled states of two quantum systems, the
qudits break the local realism stronger than the
qubits\cite{KaszlikowskiGnaciZukowskiMiklaszewskiZeilinger}, and
the entangled qudits are less influenced by  noise than the
entangled qubits. Using entangled qutrits or qudits instead of
qubits is more protective from interception. From a practical
point of view, it is clear that generating and saving the
entanglement in a controlled manner is the primary problem for the
realization of quantum computers. Maximally entangled states are
best suited for the protocols of quantum
teleportation and quantum cryptography.\\
 The entanglement and the
symmetry are two basic notions of  quantum mechanics. We study the
dynamics of multipartite systems, which are invariant at any
subsystem permutation. The aim of this work is finding exact
solutions for the dynamics of coupled qudits interacting with an
alternating magnetic field as well as the comparative analysis of
the entanglement measures in a finite chain of
 coupled qudits.\\
 The paper is organized as following. The
Hamiltonian of the anisotropic qutrit in an arbitrary alternating
magnetic field is described in Sec. II. Then the system of
equations for the description of the qutrit dynamics is derived in
the Bloch vector representation. We introduce a consistent
magnetic field, which describes an entire class of field forms. In
section III we find an analytical solution for the density matrix
in the case of isotropic interaction. Analytical  formulae, which
describe the entanglement in finite spin chains of qutrits or
quartits, are presented in Sec. IV. The results are demonstrated
graphically in Sec. V at specific parameters. The conclusions are
given in Sec. VI.
\section{Qutrit}\label{qutrit}
\subsection{Qutrit Hamiltonian and Liouville-von Neumann equation}\label{Qutrit Hamiltonian}
 We take the qutrit Hamiltonian (for the spin
particle with s=1) in the space of one qutrit $\mathrm{C}^{3}$ in the basis $%
|1>=(1,0,0),\;|0>=(0,1,0),\;|-1>=(0,0,1)$, in an external magnetic
field $\overrightarrow{{h}}=(h_{1},h_{2},h_{3})$ with anisotropy,
in the form
\begin{equation}\label{eq:1}
\hat{H}(\overrightarrow{h})
=h_{1}S_{1}+h_{2}S_{2}+h_{3}S_{3}+Q(S_{3}^{2}-\frac{s(s+1)}{3}E_{2s+1
\times 2s+1})+d(S_{1}^{2}-S_{2}^{2}),
\end{equation}
where $h_{1},\;h_{2},\;h_{3}$ are the Cartesian components of the
external
magnetic field in  frequency units (we assume $\hbar =1$, Bohr magneton $\mu_B=1$); $%
S_{1},\;S_{2},\;S_{3}$ are the spin-1
matrices\cite{finitequtritchain}; $E_{2s+1 \times 2s+1}$ is the
unity matrix; \thinspace $Q,\;d$ are the anisotropy constants.
When the constants $Q,\;d$ are zeros, then the two Hamiltonian
eigenvalues
are symmetrically placed with respect to the zero level.\\
 There exist many useful bases\cite{BertmannKrammer}.
  Allard and Hard\cite{AllardHard}(AH) formed the Hermitian basis
  $C_\alpha$
  from the linear combinations of the irreducible tensor operators.
  This basis is normalized so that
  $S_1=C_{1,x}, S_2=C_{1,y}, S_3=C_{1,z},$
 irrespective of the spin quantum number $s$. It is convenient  to
construct the spin Hamiltonian for any spin. Hereinafter we use
the Hermitian basis\cite{AllardHard}. From the physical point of
view, for important physical applications the
basis\cite{AllardHard,JMPIvanchenko} is preferred. It is not
necessary for the basis to be Hermitian since the results of the
calculations are independent of the choice of base, but there is a
significant advantage of the Hermitian basis. It is useful that
the Liouville-von Neuman equation does not involve complex numbers
and can be solved using real algebra. It makes numerical
calculations faster and simplifies the interpretation of the
equation system. The transition matrix determines the coupling
between the generalized Gell-Mann and (AH) Hermitian matrix bases.
This coupling for qutrit is presented in
Ref. 9.\\
 \indent The qutrit dynamics in a magnetic field is
described in the density matrix formalism using the Liouville-von
Neumann equation
\begin{equation}\label{eq:2}
i\partial _{t}\rho =[\hat{H},\,\rho ],~\rho (t=0)=\rho _{0}.
\end{equation}
It is convenient to rewrite Eq.~(\ref{eq:2})
 presenting the density matrix $%
\rho $ in the decomposition with a full set\cite{AllardHard} of
orthogonal Hermitian matrices $C_{\alpha }$ (further the summation
over the Greek indices will be from 0 to 8 and over the Latin ones
from 1 to 8)
\begin{equation} \label{eq:3}
\rho =\frac{1}{\sqrt{6}}C_{\alpha }R_{\alpha }.
\end{equation}
Since $\mathrm{Tr\,}C_{i}=0$ for $1\leq i\leq 8$, then from the
condition $\mathrm{Tr\,}\rho =R_{0}$ it follows that $R_{0}=1$.
And although the results are independent of the basis choice, in
this basis the functions $R_{i}=\mathrm{Tr\,}\rho \,C_{i}$ have
the concrete physical meaning\cite{AllardHard}. The values
$R_{1},R_{2},R_{3}$ are the polarization vector
Cartesian components; $R_{4}$ is the two-quantum coherence contribution in $%
R_{2}$; $R_{5}$ is the one-quantum anti-phase coherence
contribution in $R_{2}$; $R_{6}$ is the contribution of the
rotation between the phase and anti-phase one-quantum coherence;
$R_{7}$ is the one-quantum anti-phase coherence contribution in
$R_{1}$; $R_{8}$
is the two-quantum coherence contribution in $R_{1}$.\\
The Liouville-von Neumann equation in terms of the functions
$R_{i}$ takes the form of a closed system of 8 real differential
first-order equations. This system of equations can be written in
a compact form as following\cite{JMPIvanchenko,Elgin}:
\begin{equation}\label{eq:5}
\partial _{t}R_{l}=e_{ijl}h_{i}R_{j},
\end{equation}
where  $e_{ijl}$ are the  structure constants,
$h_{i}=2(h_{1},h_{2},h_{3},0,0,\frac{Q}{\sqrt{3}},0,d)$ are the
Hamiltonian components Eq.~(\ref{eq:1}) in the basis $C_\alpha$.
\subsection{The consistent field}\label{The consistent field}
Let us consider the qutrit dynamics in an alternating field of the
form
\begin{equation}\label{eq:6}
\vec{h}(t)=\left( \omega _{1}\mathrm{cn}(\omega t|k),\;\omega
_{1}\mathrm{sn} (\omega t|k),\;\omega _{0}\mathrm{dn}(\omega
t|k)\right) ,
\end{equation}
where $\mathrm{cn},\mathrm{sn},\mathrm{dn}$ are the Jacobi
elliptic functions\cite{AbramovitzStegun}. Such field modulation
under the changing of the elliptic modulus $k$ from 0 to 1
describes the whole class of field forms from
trigonometric\cite{IIRabi} ($\mathrm{cn}(\omega
t|0)=\mathrm{cos}\omega t,\;\mathrm{sn}(\omega
t|0)=\mathrm{sin}\omega t,\;\mathrm{dn}(\omega t|0)=1$ )  to the
exponentially impulse ones ($\mathrm{cn}(\omega t|1)=
\frac{1}{\mathrm{ch}\omega t},\;\mathrm{sn}(\omega
t|1)=\mathrm{th}\omega t,\;\mathrm{dn}(\omega
t|1)=\frac{1}{\mathrm{ch}\omega t}$)\cite{BambiniBerman}. The
elliptic functions $\mathrm{cn}(\omega t|k)$ and$\;
\mathrm{sn}(\omega t|k)$ have the real period $\frac{4K}{\omega
}$, while the function $\mathrm{dn}(\omega t|k)$ has a period of
half the duration. Here $K$ is the full elliptic integral of the
first kind\cite{AbramovitzStegun}. In other words, even though the
field is periodic with a common real period $\frac{4K}{\omega }$,
but as we can see, the frequency of the longitudinal field
amplitude modulation is twice as high as that of the transverse
field. We call such field consistent.\\
\indent Let us make use of the substitution $ \rho =\alpha
_{1}^{-1}r\alpha _{1}$ with the diagonal matrix $\alpha
_{1}=\mathrm{diag\,}(f, 1, f^{-1})$, where $f(\omega
t|k)=\mathrm{cn}(\omega t|k)+i\mathrm{sn}(\omega t|k).$ Then we
obtain the equation for the matrix $r$ in the form
\begin{equation}\label{eq:7}
i\partial _{t}r=[\alpha _{1}\hat{H}\alpha _{1}^{-1}-i\alpha
_{1}\partial _{t}(\alpha _{1}^{-1}),r].
\end{equation}
 The equation for the matrix $r$ without taking into account
the anisotropy can be written as following
\begin{equation}\label{eq:8}
i\partial _{t}r=[\omega _{1}S_{1}+\delta \,\mathrm{dn}(\omega
t|k)S_{3},r],\;r(t=0)=\rho _{0},\;\delta =\omega _{0}-\omega .
\end{equation}
At $k=0$ equation (\ref{eq:8})  describes the dynamics of the
qutrit in a circularly polarized
field\cite{IIRabi,MillerSuitsGarroway,GrifoniHanggi}. The exact
solutions of this equation are known, and under certain initial
conditions the explicit formulae are given in
Ref.~\refcite{NathSenGangopadhyay}. At the exact resonance,
$\omega =\omega _{0}$ it is straightforward to present
(\ref{eq:2}) in the deformed field ($k\neq0$) (\ref{eq:6})  for
the given initial condition $\rho =\rho _{0}$:
\begin{equation}\label{eq:9}
\rho (t)=\alpha _{1}^{-1}e^{-i\omega _{1}t S_{1}}\rho
_{0}e^{i\omega _{1}t S_{1}}\alpha _{1}.
\end{equation}
Explicit solutions for specific initial conditions are given in
Ref.~\refcite{finitequtritchain}.
\section{Bi-qutrit}\label{Biqutrit}
In the space $\mathrm{C}^{3}\otimes \mathrm{C}^{3}$ the bi-qutrit
density matrix can be written in the Bloch representation
\begin{equation}\label{eq:10}
\varrho =\frac{1}{6}R_{\alpha \beta }C_{\alpha }\otimes C_{\beta
},\;R_{00}=1,\;\varrho (t=0)=\varrho _{0},
\end{equation}
where $\otimes $ denotes the direct product. The functions
$R_{m0},R_{0m}$ characterize the individual qutrits and functions
$R_{mn}$ characterize their correlations.
\\
Let us consider the Hamiltonian of the system of two qutrits with
anisotropic and exchange interaction in a magnetic field in the
following form
\begin{equation}  \label{eq:11}
H_{2} =\hat{H}(\overrightarrow{h})\otimes E_{2s+1 \times 2s+1}+
E_{2s+1 \times 2s+1}\otimes
\hat{H}(\overrightarrow{\bar{h}})+JS_{i}\otimes S_{i},
\end{equation}
where $\overrightarrow{h}$ and $\overrightarrow{\bar{h}}$ are the
magnetic field vectors in frequency units, which operate on the
first and the second qutrits respectively, and $J$ is the constant
of isotropic exchange interaction.
 We study the dynamics of two qutrits in the consistent
magnetic field $\overrightarrow{h }=(\omega _{1}\mathrm{cn}(\omega
t|k)),\;\omega _{1}\mathrm{sn}(\omega t|k),\;\omega
_{0}\mathrm{dn}(\omega t|k))$,$\overrightarrow{\bar{h}}=(\varpi
_{1}\mathrm{cn}(\varpi t|k),\;\varpi _{1}\mathrm{sn}(\omega
t|k),\;\varpi _{0}\mathrm{dn}(\omega t|k))$ at the anisotropy
constants equal to 0. Let us transform the matrix density $\varrho
=\alpha _{2}^{-1}r_{2}\alpha _{2}$ with the matrix $\alpha
_{2}=\alpha _{1}\otimes \alpha _{1}$. The equation for the matrix
$r_{2}$ takes the form $i\partial
_{t}r_{2}=[\widetilde{H}(\mathrm{dn}(\omega t|k)),r_{2}]$ with the
transformed Hamiltonian $\widetilde{H}(\mathrm{dn}(\omega
t|k))$\cite{finitequtritchain}.
 \newline Since $\mathrm{dn}(\omega t|k)|_{k=0}=1$, then the transformed
Hamiltonian $\widetilde{H}$ does not depend on time, and the
solution for the density matrix in the circularly polarized field
has the form
\begin{equation}\label{eq:15}
\varrho (t)=\alpha _{2}^{-1}e^{-i\widetilde{H}t}\varrho
_{0}e^{i\widetilde{H} t}\alpha _{2}|_{k=0}.
\end{equation}%
\indent In the consistent field at resonance $\omega =\varpi
_{0}=\omega _{0}=h$ at equal $\varpi _{1}=\omega _{1}$ the
Hamiltonian eigenvalues equal to $ -2J,-J,J,J-2\omega
_{1},-J-\omega _{1},J-\omega _{1},-J+\omega _{1},J+\omega
_{1},J+2\omega _{1}$. This allows to find the exact solution in
the closed form for any initial condition since the matrix
exponent $e^{i\widetilde{H} t}$ in this case can be calculated
analytically.\newline For a larger number of the qudits with a
pairwise isotropic interaction, the generalization is evident. In
the case of interaction of qudits with a different dimensionality,
the reduction of the original system to the system with constant
coefficients can be done by choosing, for example, the
transformation matrix for spin-3/2 and spin-2 in the form%
\begin{equation}\label{eq:16}
\mathrm{diag\,}( f^{3/2},\,f^{1/2},\,f^{-1/2},\,f^{-3/2}) \otimes
\mathrm{diag\,} (f^{2},\,f,\,1,\,f^{-1},\,f^{-2}).
\end{equation}
However, the Hamiltonian eigenvalues cannot be found in a simple
analytical form because of the lowering of the system's symmetry.
\section{Analytical formulae for entanglement measures}\label{Entanglement in the qutrits}
\subsection{Entanglement in the bi-qutrit}\label{Entanglement in the bi-qutrit}
 For the initial maximally
entangled state which is symmetrical at the particle permutation
\begin{equation}\label{eq:17}
|\psi >=\frac{1}{\sqrt{3}}\sum_{i=-1}^{1}|i>\otimes |i>,
\end{equation}
in the consistent field at the resonance $\omega =\varpi
_{0}=\omega _{0}=h$ at equal $\varpi _{1}=\omega _{1},$ the exact
solution for the correlation functions is given in
Ref.~\refcite{finitequtritchain}. The correlation functions have
the property $R_{\alpha \beta }=R_{\beta \alpha }$, i.e. the
symmetry is conserved during the evolution, since the initial
state and the Hamiltonian are symmetric with respect to the
particle permutation.\newline Given the exact solution, one can
find the negative eigenvalues of the partly transposed matrix
$\varrho ^{pt}=(T\otimes E)\varrho $\ (here $T$ denotes the
transposition): $\epsilon _{1}=\epsilon
_{2}=-\frac{1}{27}\sqrt{69+28\cos 3Jt-16\cos 6Jt} ,\;\epsilon
_{3}=-\frac{1}{27}\left( 5+4\cos 3Jt\right)$. The absolute value
of the sum of these eigenvalues
\begin{equation}
m_{VW}=\sum_{i=1}^{3}| \epsilon _{i}|
\end{equation}
 defines the
entanglement measure (negativity) between the qutrits\cite{VidalWerner}.\\
 The entanglement between the qudits can be
described quantitatively with the measure\cite{SchlienzMahler}
\begin{equation}\label{eq:19}
m_{SM}=\sqrt{\frac{1}{D-1}(R_{ij}-R_{i0}R_{0j})^{2}},
\end{equation}
where D is the basis dimension (for qutrit $D=9$). This measure
equals to 0 for the separable state and to 1 for the maximally
entangled state, and it is applicable for both pure and
mixed states.\\
 That is why for the maximally entangled initial
state of two qutrits, the entanglement in the consistent field is
defined by the formulae with the found solution for the density
matrix
\begin{equation}\label{eq:20}
m_{SM}=\frac{1}{81}\sqrt{4457+2776\cos 3Jt-632\cos 6Jt-56\cos
9Jt+16\cos 12Jt}.
\end{equation}
This measure is numerically equivalent to the measure
$m_{VW}$\cite{VidalWerner,PermutaionalSymTothGuhne} which is
defined by the absolute value of the sum of the negative
eigenvalues.
\\
 According to the
definition for $N$-qudit pure state\cite{PanLiuLuDraayer}, the
entanglement measure equals to
\begin{equation} \label{eq:21}
\eta _{N}=\frac{1}{N}\sum_{i=1}^{N}S_{i},
\end{equation}%
where $S_{i}=-\mathrm{Tr\,}\rho _{i}\log _{b}\rho _{i}$ is the
reduced von Neumann entropy, the index $i$ numerates the
particles, i.e. the other particles are traced out. We use the
logarithm to the base $b$ to ensure that the maximal measure is
normalized to 1.
The base $b$ equals to 3 in the qutrit case.\\
 Since
the qutrit reduced matrix eigenvalues equal to $\lambda
_{1}=\lambda _{2}=\frac{1}{27}(5+4\cos 3Jt),\;\lambda
_{3}=\frac{1}{27}(17-8\cos 3Jt),$ then the entanglement measure in
the bi-qutrit takes the form
\begin{equation}\label{eq:22}
\eta _{2}=-\sum_{i=1}^{3}\lambda _{i}\log _{3}\lambda _{i}.
\end{equation}
Normalized to unity the measure I-concurrence which is easy to
calculate
 is defined by the formulae\cite{Mintert}
\begin{equation}\label{eq:23}
m_{I}=\sqrt{\frac{d}{d-1}}\sqrt{(1-\mathrm{Tr\,}\rho _{1}^{2})}
=\frac{1}{9} \sqrt{57+32\cos 3Jt-8\cos 6Jt},
\end{equation}
where $d=3$ for a qutrit,\; $\rho _{1}=\frac{1}{\sqrt{6}}C_{\alpha
}R_{\alpha 0}$ is the
reduced qutrit matrix.\\
The measures $m_{VW},\;m_{SM},\;\eta _{2},\;m_{I}$ do not depend
on the parameters of the consistent field, the sign of the
exchange constant at zero anisotropy parameters. It should be
noted that the Wootters entanglement measure (the concurrence) in
the system of two qubits with an isotropic interaction in a
circularly polarized field at resonance is also independent of the
alternating field
amplitude\cite{ShiXunZhangQinShengZhuXioYuKuang}, but depends on
the exchange constant $J$ and the initial conditions only.\\
\indent The numerical solution of the Liouville-von Neumann
equation shows that if an identical external field operates on
every qudit, the free Hamiltonian and the interaction  Hamiltonian
are commutative operators, the measures considered are determined
only by the symmetric two-body interaction with the interaction
constant of  $J$. If a different field operates on every qudit,
there arises a broken permutation symmetry of the total
Hamiltonian, which changes  the entanglement dynamics. Thus it is
possible to control entanglement by changing the
parameters of an external field.\\
 \indent It is possible to show that the distance
measure\cite{VedralPlenio,Horodecki}  $\sqrt{{\rm
Tr\,}(\rho(t)-\rho_0)^2}$ depends on
 the parameters of the consistent field.\\
 At a zero external field the entanglement measure (\ref{eq:19})~
takes the analytical form at equal non-zero anisotropy parameters
$Q=d=\overline{d}= \overline{Q}$
\begin{equation}\label{eq:24}
m_{SM}(Q)=\frac{1}{\left( 9J^{2}+8QJ+16Q^{2}\right) ^{2}}\sqrt{
\sum_{k=0}^{4}q_{k}\cos \left(
k\sqrt{9J^{2}+8QJ+16Q^{2}}\,t\right)},
\end{equation}
where $
q_{0}=4457J^{8}+11616QJ^{7}+47392Q^{2}J^{6}+85888Q^{3}J^{5}+163072Q^{4}J^{4}
+194560Q^{5}J^{3}+221184Q^{6}J^{2}+131072Q^{7}J+65536Q^{8}
$;\\$q_{1}=8J^{2}(J+2Q)^{2}\left(
347J^{4}+518QJ^{3}+1440Q^{2}J^{2}+1504Q^{3}J+1024Q^{4}\right) $;\\
$q_{2}=-8J^{2}(J+2Q)^{2}\left(
79J^{4}+76QJ^{3}+320Q^{2}J^{2}+448Q^{3}J+256Q^{4}\right) $;\\
$q_{3}=-8J^{3}(7J-4Q)(J+2Q)^{3}(J+4Q)$,\
$q_{4}=16J^{4}(J+2Q)^{4}$.
\subsection{Entanglement in the chain of qutrits}\label
{Entanglement in the chain of qutrits}
We consider the Hamiltonian of the chain of $N$ qutrits with the
pairwise isotropic interaction in the consistent field $\vec{h}(t)
 $ (\ref{eq:6}) at resonance in the following form
\begin{equation}\label{eq:25}
H_{N}=\sum (\vec{h}(t)\overrightarrow{S}\otimes \overbrace{
E\otimes \dots \otimes E}^{N-1}+J\overrightarrow{S}\otimes
\overrightarrow{S} \otimes \overbrace{E\otimes \dots \otimes
E}^{N-2}),
\end{equation}%
where the summation is over different possible positions of
$\overrightarrow{ S}$ in the direct products. Because the
maximally entangled state of $N$
qutrits%
\begin{equation}\label{eq:26}
|\phi >_N=\frac{1}{\sqrt{3}}\sum_{i=-1}^{1}|i>^{\otimes N}
\end{equation}
and the Hamiltonian (\ref{eq:25}) have a permutation symmetry, it
follows that the density matrix of $N$ qutrits has
symmetric correlation functions.

\indent The entanglement measures for many-particle multi-level
quantum systems have not been studied enough and are difficult to
calculate in the analytical form, that is why we will present
 analytical formulae only for the entropy measure $\eta _{N}$, which
is defined by the eigenvalues of the reduced one-particle matrices
for each qutrit. As the result of the mentioned symmetry the
reduced matrices are equal to each other. Therefore the
entanglement measure for $N$ qutrits reads
\begin{equation}\label{eq:27}
\eta _{N}=-\sum_{i=1}^{3}r_{i}\log _{3}{r_{i}}.
\end{equation}
The eigenvalues of the reduced matrices for 3, 4, 5, and 6 qutrits
are presented in the table below
\begin{equation}\label{eq:28}
\begin{array}{ccc}
N\setminus r_{i} & r_{1}=r_{2} & r_{3} \\
&  &  \\
3 & \frac{29-4\cos 5Jt}{75} & \frac{17+8\cos 5Jt}{75} \\
4 & \frac{905-98\cos 3Jt-72\cos 7Jt}{2205} & \frac{395+196\cos
3Jt+144\cos
7Jt}{2205} \\
5 & \frac{16919-1944\cos 5Jt-800\cos 9Jt}{42525} &
\frac{8687+3888\cos
5Jt+1600\cos 9Jt}{42525} \\
6 & \frac{21977-1694\cos 3Jt-1936\cos 7Jt-560\cos 11Jt}{53361} &
\frac{
9407+3388\cos 3Jt+3872\cos 7Jt+1120\cos 11Jt}{53361}. \\
&  &
\end{array}%
\end{equation}
The \ measures $\eta _{3},\;\eta _{4},\;\eta _{5},\;\eta _{6}$
 do not depend on the sign of the exchange constant like the
measure $\eta _{2}$.
\subsection{Entanglement in the bi-quartit}\label{Entanglement in bi-quartit}
The applied approach for qutrits is translated to qudits. For a
spin-3/2 particle or a four-level system, also denoted as a
quartit, we take the Hamiltonian in the space $\mathrm{C}^{4}$ in
the basis $
|3/2>=(1,0,0,0),\;|1/2>=(0,1,0,0),\;|-1/2>=(0,0,1,0),\;|-3/2>=(0,0,0,1)$
and use the matrix representation  of a complete set of the
Hermitian orthogonal operators.\\
 We will find
the analytical formulae in the bi-quartit and in the 3 quartits
(in bi-pentit, see below) with a pairwise isotropic interaction of
the initial maximally entangled state in a consistent magnetic
field at resonance without taking into account the anisotropy.

  The
negative eigenvalues of the partly transposed matrix $\varrho
^{pt}$
are equal to  $ \lambda _{1}=\frac{1}{100} (-13-12 \cos 5 J t),\,
\lambda _{2}=\lambda _{3}=\lambda _{4}=\lambda _{5}=-\frac{1}{100}
\sqrt{409+288 \cos 5 J t-72 \cos 10 J
   t},\,\lambda _{6}=\frac{1}{100} (-37+12 \cos 5 J t).$
   The  entanglement measure in the bi-quartit  equals
\begin{equation}\label{eq:29}
 m_{VW}^{bi-qrt}=\sum_{i=1}^{6}|\lambda _{i}|.
\end{equation}
The entanglement between the quartits is described quantitatively
with the measure\cite{SchlienzMahler}
\begin{equation}
m_{SM}^{bi-qrt} =
 \frac{
 \sqrt{1803365+191616
\cos 5 J t-35808
   \cos 10 J t-6912 \cos 15 J t+864 \cos 20J t}}{625 \sqrt{5}}.
\end{equation}
 Since the quartit reduced matrix eigenvalues equal to $
\lambda_1=\lambda_2=\frac{1}{100}(13+12 \cos 5 J t)
,\,\lambda_3=\lambda_4=\frac{1}{100}(37-12 \cos 5 J t) $, hence
the measure $\eta _{2}$ reads
\begin{equation}\label{eq:30}
\eta _{2}^{bi-qrt}=-\sum_{i=1}^{4}\lambda _{i}\log _{4}\lambda
_{i}.
\end{equation}
The  I-concurrence is equal to
\begin{equation}\label{eq:31}
m_{I}^{bi-qrt}= \frac{1}{25} \sqrt{553+96 \cos 5 J t-24 \cos 10 J
t}.
\end{equation}
\subsection{Entanglement in the quartit chain}\label
{Entanglement in the quartit chain}
\indent The eigenvalues of the reduced matrices for 3 quartits are
equal to  $ r_1=r_2=0.141+0.068 \cos \frac{5 J t}{2}+0.04 \cos 8 J
t,\, r_3=r_4=0.359-0.068 \cos \frac{5 J t}{2}-0.04 \cos 8 J t$.
Therefore the entanglement measure for $3$ quartits is following
\begin{equation}\label{eq:32}
\eta _{3}^{qrt}=-\sum_{i=1}^{4}r_{i}\log _{4}{r_{i}}.
\end{equation}
\subsection{Entanglement in the bi-pentit}\label{Entanglement in bi-pentit}
For the spin-2 particle or a 5-level system, also denoted as a
pentit, we take the  Hamiltonian  in the space $\mathrm{C}^{5}$ in
the basis $ |2>=(1,0,0,0,0),\;|1>=(0,1,0,0,0),\;|0>=(0,0,1,0,0),
\;|-1>=(0,0,0,1,0),\;|-2>=(0,0,0,0,1).$
\\
 \indent The entanglement between the
pentits is described using the measure\cite{SchlienzMahler}
$m_{SM}^{bi-pnt}= (0.802+0.106 \cos 3 J t-0.019 \cos 4 J t+0.242
\cos 7 J
   t-0.098 \cos 10 J t-0.088 \cos 14 J t+0.067 \cos 17 J t-0.014 \cos 20 J
   t)^{1/2}$.  The  I-concurrence is determined by the formulae $m_{I}^{bi-pnt}=
    (0.791 + 0.114 \cos 3 J t - 0.018 \cos 4 J t - 0.005 \cos 6 J t +
    0.230 \cos 7 J t - 0.079 \cos 10 J t - 0.079 \cos 14 J t +
    0.060 \cos 17 J t - 0.015 \cos 20 J t)^{1/2}$. We have replaced
   the exact bulky rational coefficients by its decimal approximations
   and the terms less than 0.001 have  removed for inconvenience
   reduction.
\\
 The pentit reduced matrix eigenvalues are equal to $
p_1=p_2=\frac{1}{6125}(1173-140 \cos 3 J t+640 \cos 7 J t-448 \cos
10 J t),\,p_3=p_4=\frac{1}{6125}(513+280 \cos 3 J t+320 \cos 7 J
t+112 \cos 10 J t),\,p_5=\frac{1}{6125}(2753-280 \cos 3 J t-1920
\cos 7 J t+672 \cos 10 J t) $, hence the measure $\eta _{2}$ reads
\begin{equation}\label{eq:33}
\eta _{2}^{bi-pnt}=-\sum_{i=1}^{5}p_{i}\log _{5}p_{i}.
\end{equation}
\\
\indent  All the measures
 do not depend on the sign of the exchange constant and the parameters
  of the consistent field at zero anisotropy parameters.
\begin {figure}
 \includegraphics[width=0.35\textwidth]{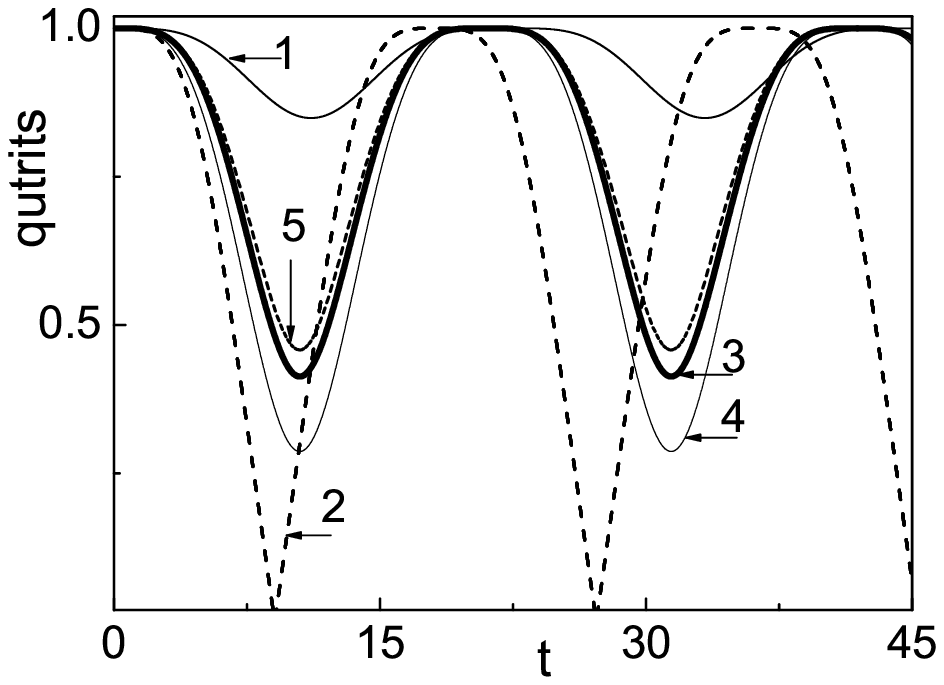}
  \hfill
  \includegraphics [width=0.35\textwidth]{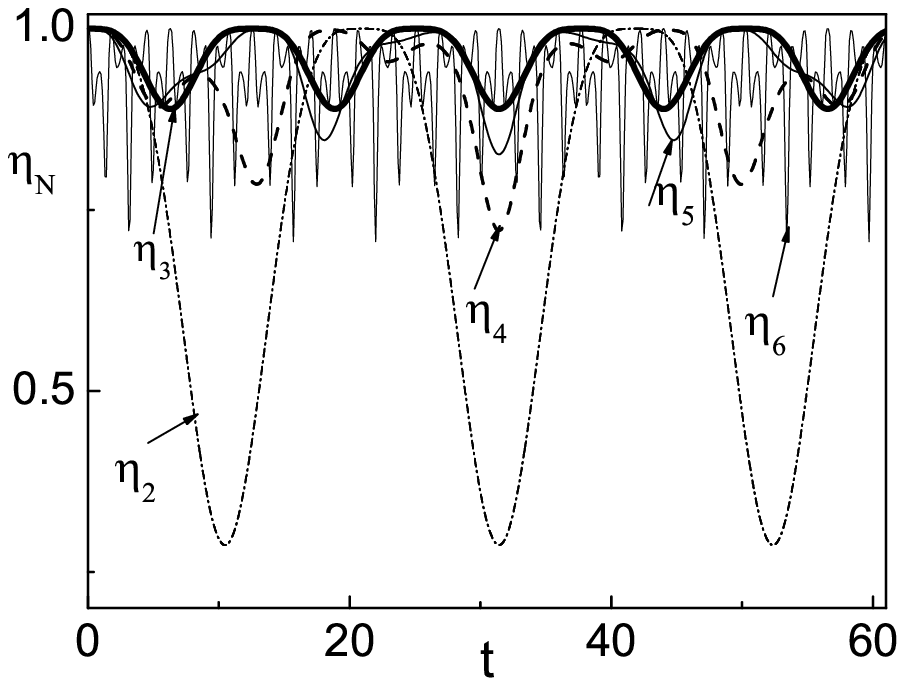}
 \\
 \parbox[t]{0.45\textwidth}{
\caption{\label{EntMSJmJpEntropyFig1} Dynamics in the bi-qutrit:
in the zero external field at equal anisotropy constants $
Q=0.0250,\,J=-0.1$ (curve 1) and  $J=0.1$ (curve 2);  \thinspace
in the consistent field at $J=0.1, Q=0$ the curve 3 shows complete
coincidence of  $m_{VW}$ and $m_{SM}$; the curve 4 is the  measure
$\eta _{2}$; $m_{I}$ is the curve 5.}} \hfill
 \parbox[t]{0.45\textwidth}
 {\caption {\label{h2h3h4h5h6Fig2}Disentanglement of the maximally entangled state
 in the chain of 2,\thinspace 3,\thinspace
4,\thinspace 5,\thinspace  6 qutrits with $J=0.1$.}}
 \end {figure}
\section{Numerical results}\label{Numerical results}
\begin{figure}
 \includegraphics[width=0.35\textwidth]{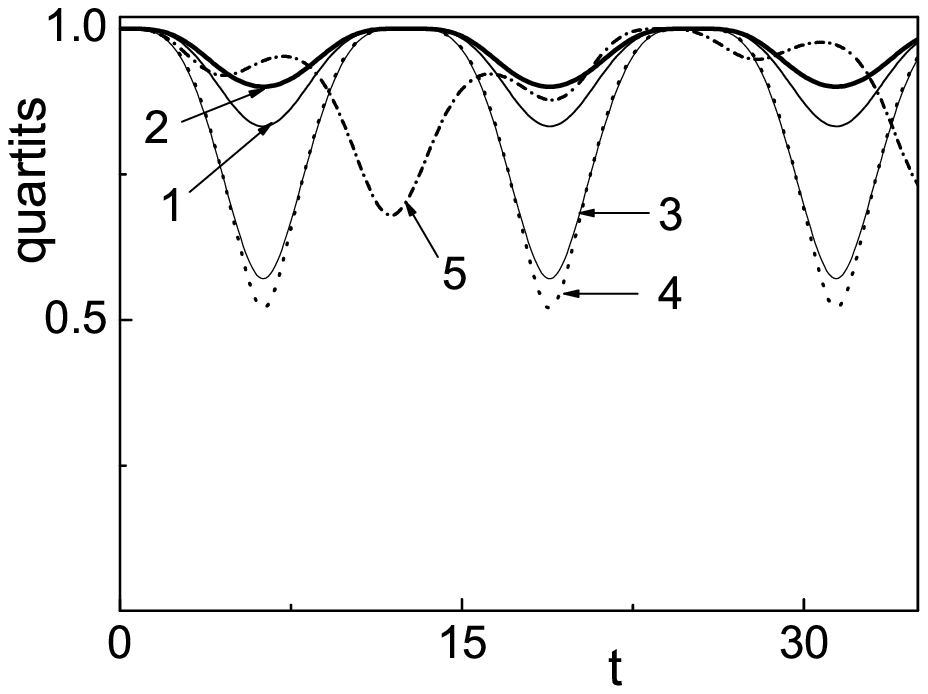}
  \hfill
  \includegraphics [width=0.35\textwidth]{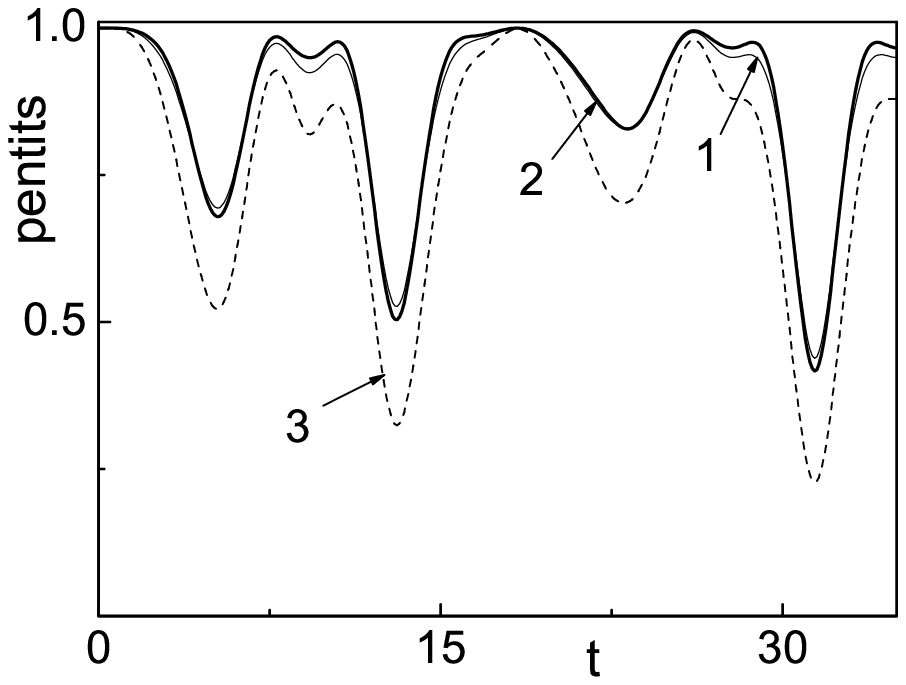}
 \\
 \parbox[t]{0.45\textwidth}{
\caption{\label{bi_quartit_measureFig3}Disentanglement of the
maximally entangled state  in the chain of 2,\thinspace 3 quartits
with $J=0.1$. The measures $m_{I}^{bi-qrt}$,\;
$m_{SM}^{bi-qrt}$,\;$\eta _{2}^{bi-qrt}$, $m_{VW}^{bi-qrt}$, are
presented by the curves 1, 2, 3, 4 respectively;  $\eta
_{3}^{qrt}$ is the curve 5.}} \hfill
 \parbox[t]{0.45\textwidth}
 {\caption {\label{bi_pentit_measureFig4}Disentanglement of the maximally
entangled state  in the chain of 2 pentits with $J=0.1$. The
measures $m_{I}^{bi-pnt}$,\; $m_{SM}^{bi-pnt}$,\;$\eta
_{2}^{bi-pnt}$,  are presented by the curves 1, 2, 3
respectively.}}
 \end {figure}

%
Although the analytical expressions for the measures in a
bi-qutrit $ m_{VW},\;m_{SM}$ are different, but the numerical
values are practically identical. The maximal deviation in the
rectangle
$(1\geq J\geq 0.01)\times (100\geq t\geq 0)$ equals $0.014$.\\
Measures $\eta _{2}$ and $m_{I}$ qualitatively coincide with the
measures $
m_{VW},\;m_{SM}$. \\
We have found that the anisotropy of the qutrits disentangles
them, namely the entanglement is decreased down to 0.001 (see
graphs 1 and 2 in Fig.\ref{EntMSJmJpEntropyFig1}).\\
In the constant longitudinal field $\overrightarrow{h}=-
\overrightarrow{\overline{h}}=(0,\,0,\,\omega _{0})$ (the
bi-qutrit Hamiltonian eigenvalues are equal to
$J,J,x_{1},x_{2},x_{3},-p,-p,p,p$, where $ x_{1},x_{2},x_{3}$ are
the roots of the equation ${
x^{3}+2x^{2}J-p^{2}x-2J^{3}=0,\,p=\sqrt{J^{2}+\omega _{0}^{2}}})$
the Hamiltonian contains the asymmetric part, thus it follows that
the density matrix for the initial symmetric state will not be
symmetric because of the breaking of the symmetry of the particle
permutations. The analytical solution is cumbersome. In the
constant longitudinal impulse field $\overrightarrow{h
}=-\overrightarrow{\overline{h} }=(0,\,0,\,2\,(\theta
((t-17)(t-60))+\theta ((40-t)(57-t)(t-60))))$ the entanglement
dynamics is blocked\cite{finitequtritchain} at $\omega _{0}\gg J$
. This points to the
possibility to control the entanglement.\\
In Fig.\ref{h2h3h4h5h6Fig2} we present the comparative dynamics of
the entropy  measure in the finite  qutrit chain. The
disentanglement dynamics of the measures $\eta _{3},\eta _{4},\eta
_{5},\eta _{6}$ is similar to the one in the case of two qutrits,
but with smaller oscillation amplitude, i.e. larger number of the
qutrits disentangles less than two qutrits ($0.889\leq \eta
_{3}\leq 1$).\\
\indent The  measures in bi-quartit, as shown  in
Fig.\ref{bi_quartit_measureFig3},  qualitative coincide, almost
completely $ \eta _ {2} ^ {bi-qrt} $ and $m _ {VW} ^ {bi-qrt} $.
The disentanglement 3 quartits is insignificant
less than in bi-quartit.\\
 \indent The disentanglement measures in bi-pentit,
as shown  in Fig.\ref{bi_pentit_measureFig4}, qualitative
coincide, almost completely $m_{I}^{bi-pnt}$ and $m _ {SM} ^
{bi-pnt} $.
\section{Conclusion}\label{Conclusion}
 The comparative analysis of the bi-qutrit
entanglement measures on the base of the analytical solution for
the density matrix demonstrates that, in spite of the different
approaches to the derivation of the formulae for the entanglement,
all the formulae yield quite close results (Fig.
\ref{EntMSJmJpEntropyFig1}), and the measures $m_{VW}\
$and$\;m_{SM}$ are practically equal. This is in accordance with
the general results for the entanglement in the systems with a
permutational
symmetry\cite{PermutaionalSymTothGuhne}.\\
The analytical formulae for the measures $\eta _{3},\eta _{4},\eta
_{5},\eta _{6}$ are similar to the measure for two qutrits $\eta
_{2}$, but with a numerically smaller oscillation amplitude, i.e.
the larger number
of the qutrits disentangles fewer than two qutrits.\\
\indent Nevertheless, the comparison of measures in two coupled
qutrits, quartits, and pentits
\begin{equation}\label{eq:34}
\begin{array}{cc}
 bi- qutrit & m_{VW} \cong m_{SM}\\
 bi- quartit & \eta _{2}^{bi-qrt} \cong m_{VW}^{bi-qrt}\\
 bi- pentit & m_{SM}^{bi-pnt} \cong m_{I}^{bi-pnt}\\
\end{array}
\end{equation}
on the base of analytical solutions shows the absence of a full
coincidence of the measures even in a particular case of
disentangling a maximally entangled state. In other words, it is
impossible to prefer any measure,
 there remains therefore the question concerning the
 quantitative determination of
entanglement even in case of two multi-level particles.\\
%
%
 The author is grateful to A.~A. Zippa   constant invaluable support.

\end{document}